# Strain control of real-and lattice-spin currents in a silicene junction

Sarayut Phonapha[a], Assanai Suwanvarangkoon[a], Bumned Soodchomshom[a,*]

[a]Department of Physics, Faculty of Science, Kasetsart University, Bangkok 10900, Thailand

*Corresponding author

Email address: Bumned@hotmail.com; fscibns@ku.ac.th

**Abstract**

We investigate real- and lattice-spin currents controlled by strain in a silicene-based junction, where chemical potential, perpendicular electric field and circularly polarized light are applied into the strained barrier. We find that the junction yields strain filtering effect with perfect strain control of real- (or lattice-) spin currents. (i) By applying electric field without circularly polarized light we show that total current is carried by pure lattice-spin up (or down) electrons tunable by strain. (ii) When circularly polarized light is irradiated onto silicene sheet without applying electric field, total current is carried by pure real-spin up (or down) electrons tunable by strain. High conductance peaks associated with pure real-(or lattice-) spin currents in case ii(or i) occur at specific magnitude of strain, yielding strain filtering effect. Magnitudes of filtered strain due to pure real- (or lattice-) spin currents may be tunable by varying chemical potential. Sensitivity may be enhanced by increasing thickness of strained barrier. Significantly, (iii) when both perpendicular electric field and circularly polarized light are applied, the total current is carried by three species of electron groups tunable by strain. This may lead to controllable numbers of electron species to transport. This result shows that strain filtering effect in a silicene-based junction is quite different from that in graphene junction. Our work reveals potential of silicene as a nano-electro-mechanical device and spin-valleytronic applications.

## 1. Introduction

Silicene, a silicon-based hexagonal lattice akin to graphene, has recently been drawn much attention [1-3], since large-area silicene sheet on Ag (111) has epitaxially been synthesized in the laboratory [4]. Field effect transistor using silicene on Ag (111) operated at room temperature [5] is one of silicene-based applications [6-8] that guarantees it as a promising material for nanoelectronics. Unlike graphene, spin-orbit interaction (SOI) and buckled atomic structure lead to spin-valley-dependent energy gap tunable by perpendicular electric field in silicene [8, 9]. Silicene exhibits 2-dimensional topological insulators [10, 11] due to the presence of intrinsic spin orbit interaction. Similar to graphene, the carriers in silicene possess Dirac fermions due to its hexagonal lattice. The difference is that at the Dirac points, mass of Dirac fermions in silicene is valley-dependent [12]. Under the perpendicular electric and exchange fields, the currents in silicene junctions are spin-valley dependent [8, 13-20]. Perfect spin-valley polarization controlled by electric field has been predicted in a sublattice-dependent ferromagnetic silicene-based junction [8]. Such behavior shows high potential of silicene in spintronics and valleytronics applications. Recently, silicene atomic structure under mechanical strain has been studied by several authors [21-25]. Silicene can sustain uniaxially armchair- or zigzag-direction strain up to about 15 % [21, 22]. Under large applying strain, the carrier behaves like anisotropic Dirac fermions [24]. Effect of biaxial strain on electronic properties in silicene was also investigated [25].

Important for applications of electro-mechanical devices, spin and valley currents in silicene based junction under the influence of strain have been investigated [26, 27]. Similar to graphene, strain in silicene may behave like pseudo vector potentials due to distortion of honeycomb atomic structure. The effect of strain on spin and valley-dependent transmissions in silicene was studied [26]. Spin and valley transports in dual-gated silicene junction have been also studied [27]. Thermal transport in silicene due to effect of tensile strain was next investigated[28]. The thermal conductivity of silicene is found to be about 20% of that in bulk silicon. Recently, a possible control of **pseudospin** or **lattice-spin** currents in ferromagnetic silicene has been proposed [29]. Physically, lattice-spin up(down) state is equivalent to wave function of electron at A(B) sublattice. Unequal currents between A and B sublattices lead to lattice-spin polarized currents. It can be said that the currents in silicene are split into two groups in real space. One moves in A-sublattice, while the other moves in B-sublattice. The presence of spin orbit interaction in silicene leads to control of such currents [29]. Pristine graphene lacks this property due to weak spin orbit interaction in its atomic structure [30-32]. Recently, topological phase transition induced by circularly polarized photon in silicene has been proposed [10, 33]. Interaction of circularly polarized photon may induce energy gap to change topological phase. Spin-valley current filtering effect in silicene under circularly polarized photon irradiating into the barrier, was investigated [19, 34].

In this letter, we investigate real-and lattice-spin conductances using Landauer formula [35] in a silicene-based NM/St/NM junction, where NM and St are equilibrium silicene region and strained silicene barrier, respectively. Chemical potential, electric field and circularly polarized light are assumed to be applied into the strained barrier [10]. A smoothly small strain is applied only in the armchair direction in the barrier region. In our model, we adopt tight-binding method to construct Hamiltonian of electrons near the Dirac points, under the influence of strain field. Uniaxially strain-induced distortion may lead to change of the magnitude of nearest hopping energy in silicene atomic structure similar to model studied in strained graphene system [36]. In our work, we would like to show perfect strain control of lattice- and real- spin currents in the junction related to perpendicular electric field and circularly polarized photon, respectively. This may be important for applications of spin valleytronics and strain sensor. This effect has not been considered in previous works [26, 27].

## 2. Theory and model

Firstly, let us consider the tight-binding model for silicene under strain in an atomic model with 3-nearest ($\vec{d}_i$) and 6-next-nearest ($\vec{d}_{ij}$) neighbor atoms as depicted in Fig.1a. In the model we may consider strain field as perturbation of hopping energy [36] and spin orbit interaction, because silicon atoms are shifted in real space. Therefore, the tight-binding Hamiltonian in the strained barrier, St-region, may be given by

$$
\begin{aligned}
H_{St} = & \sum_{<i,j>\alpha} \left( -t_0 + \delta_j \right) c_{i\alpha}^\dagger c_{j\alpha} + i \sum_{<<i,j>>\alpha\beta} \left( t_{so} + \delta_{soj} \right) \nu_{ij} c_{i\alpha}^\dagger \sigma_{\alpha\beta}^z c_{j\beta} \\
& -i \sum_{<<i,j>>\alpha\beta} \left( t_R + \delta_{Rj} \right) \mu_{ij} c_{i\alpha}^\dagger \left( \vec{\sigma} \times \hat{d}_{ij} \right)_{\alpha\beta}^z c_{j\beta} \\
& + \sum_{i\alpha} \eta_i eDE_z c_{i\alpha}^\dagger c_{i\alpha} \\
& + i \frac{\Delta_\Omega}{3\sqrt{3}} \sum_{<<i,j>>\alpha\beta} \nu_{ij} c_{i\alpha}^\dagger c_{j\beta} \\
& - \sum_{i\alpha} \mu c_{i\alpha}^\dagger c_{i\alpha} ,
\end{aligned}
\tag{1}
$$

where $c_{i\alpha}^\dagger$ creates an electron with spin polarization $\alpha$ at site "i". The notations $<i,j>$ and $<<i,j>>$ denote running over all nearest and next-nearest neighbor hopping sites, respectively. The first term represents the strain-induced perturbation of nearest-neighbor hopping energy of $-t_0 + \delta_j$ where $t_0 = 1.6$ eV [12] is hopping energy for unstrained silicene. The second term represents the strain-induced perturbation of spin orbit coupling of $t_{so} + \delta_{soj}$, where $t_{so}3\sqrt{3} = \Delta_{so} = 3.9$ meV [12] is the strength of spin orbit interaction for unstrained silicene. The notation $\vec{\sigma} = <\sigma^x, \sigma^y, \sigma^z>$ denotes Pauli spin vector acting on real-spin space, and $v_{ij} = 1(-1)$ when the next-nearest neighboring hopping is anticlockwise (clockwise). The third term represents the strain-induced perturbation of Rashba spin orbit coupling of $t_R + \delta_{Rj}$ where $(3/2)t_R = \lambda_R = 0.7$ meV [12] is Rashba spin orbit coupling for unstrained case [37]. $\mu_{ij} = 1(-1)$ for A(B) sublattice, and $\hat{d}_{ij} = \vec{d}_{ij} / |\vec{d}_{ij}|$ where $\vec{d}_{ij} = \vec{d}_i - \vec{d}_j$ for $i \neq j$. We note that spin-orbit couplings of nearest-neighbor atoms <i,j> is zero due to the structure's mirror symmetry with respective to an arbitrary bound for coupling between different sub-lattices. It is nonzero for next-nearest-neighbor atoms <<i,j>> for coupling between the same sub-lattices, represented in the second and third terms [12]. The fourth term represents perpendicular electric-field-induced potential of $\eta_i eDE_z$, where $\eta_i = 1(-1)$ for A(B) sublattice. e, D and $E_z$ are electronic charge, buckling parameter and perpendicular electric field, respectively. Breaking the inversion symmetry by electric-field-induced staggered AB sublatices potential leads to energy gap of $2eDE_z$. The fifth term represents the interaction Hamiltonian for circularly polarized photon irradiated onto silicene sheet [10]. The in-plane vector potential of circularly polarized photon is given by $\vec{A}(t) = A \langle \sin \Omega t, \cos \Omega t \rangle$ where $A$ is dimensionless intensity and $\Omega > 0 (< 0)$ for right (left) circulation. This photo-interaction leads to $\Delta_\Omega = \hbar v_F^2 A^2 \Omega^{-1}$ [10, 33]. The last term represents chemical potential $\mu$ at the barrier. In unstrained silicene, NM-region, there is no strain saying $\delta_j = \delta_{soj} = \delta_{Rj} = 0$ with electric field $E_z = 0$ and chemical potential $\mu = 0$. The tight-binding Hamiltonian in NM-regions may be reduced to the usual form [12]

$$H_{NM} = -t_0 \sum_{<ij>\alpha} c_{i\alpha}^\dagger c_{j\alpha} + it_{so} \sum_{<<ij>>\alpha\beta} v_{ij} c_{i\alpha}^\dagger \sigma_{\alpha\beta}^z c_{j\beta}$$
$$-it_R \sum_{<<ij>>\alpha\beta} \mu_{ij} c_{i\alpha}^\dagger \left( \vec{\sigma} \times \hat{d}_{ij} \right)_{\alpha\beta}^z c_{j\beta} . \qquad (2)$$

In silicene, $t_R$ is much smaller than $t_{so}$. Thus, $t_R$ may be neglected near the Dirac points [12, 37]. The present model in Fig. 1a, we assume that a small strain is applied only in the armchair direction in the St-region. We may get $\delta_2 = \delta_3 = \delta_{soj} \ll \delta_1$, while $\delta_1 = \delta$ cannot be canceled. The perturbed parameters which are not parallel to the armchair direction may be cancelled to get $\delta_2 = \delta_3 = \delta_{soj} \cong 0$. Small perturbation of hopping energy $\delta$ induced by a small uniaxial strain in honeycomb lattice may be proportional to magnitude of strain [36, 39]. We may get strain-dependent nearest-hopping energy $t(s) = -t_0 + \delta = g \times (-t_0) \cong -t_0 + \gamma s$, where $g = e^{-\gamma s} \cong 1 - \gamma s$ [36]. In silicene $\gamma$ is still an unknown parameter and "s" stands for magnitude of strain, while in graphene, $\gamma \cong 3.37$ [36]. For strain-dependent hopping energy model, the factor "g" may be smaller than one ( $g \leq 1$ ) for positive strain [36, 40], since strain causes larger distance between atoms than that in equilibrium system to reduce the magnitude of hopping energy. Using these conditions, the Hamiltonian of quasiparticle derived from eq.1 near the Dirac points can be used to describe motion of electron in the **St-region.** It may be obtained as of the form

$$\hat{h}_{St} = \eta \left( v_F \hat{p}_x + \eta \delta \right) \tau^x + v_F \hat{p}_y \tau^y + \left( \eta s \Delta_{so} + \Delta_z + \eta \Delta_\Omega \right) \tau^z - \mu, \qquad (3)$$

where $v_F = 5.5 \times 10^5$ m/s is the Fermi velocity and $\Delta_z = eDE_z$. $\eta = 1(-1)$ denotes electron in $k(k')$ valley, $s = 1(-1)$ denotes electron with real-spin $\uparrow(\downarrow)$ and $\vec{\tau} = <\tau^x, \tau^y, \tau^z>$ denotes Pauli spin metric vector acting on lattice-spin space. The momentum operators is $\hat{p}_{x(y)} = -i\hbar \partial_{x(y)}$. In the NM-regions, the Hamiltonian follows eq.2 or takes $\delta = \Delta_z = \delta = \mu = 0$ into eq.3, and is given by [10, 11, 37]

$$\hat{h}_{NM} = \eta v_F \hat{p}_x \tau^x + v_F \hat{p}_y \tau^y + \eta s \Delta_{so} \tau^z, \qquad (4)$$

The wave functions of electron in NM regions satisfy equation $\hat{h}_{NM} \left| \psi_{\eta s, \pm k_y} \right\rangle = E \left| \psi_{\eta s, \pm k_y} \right\rangle$ where the wave vector of electron is $\vec{k} = \left\langle k_{//}, \pm k_y \right\rangle$ and energy is E with conservation component $k_{//} = \sqrt{E^2 - \Delta_{so}^2} \cos \theta / \hbar v_F$ where $\theta$ denotes incident angle in the NM-regions. Using the lattice-spin operator defined as $\hat{S}_{ps} = \dfrac{\hbar}{2} \vec{\tau}$, the expectation value of lattice-pseudo spin related to state $\left| \psi_{\eta s, \pm k_y} \right\rangle$ in NM-regions is calculated via the formula $\left\langle \vec{S}_{sp} \right\rangle = \left\langle \psi_{\eta s, \pm k_y} \left| \hat{\vec{S}}_{ps} \right| \psi_{\eta s, \pm k_y} \right\rangle$, to get [29,38]

$$\left\langle \vec{S}_{sp} \right\rangle = \eta \frac{\hbar}{2} \sqrt{1 - \left( \frac{\Delta_{so}}{E} \right)^2} \left[ \cos(\eta \theta) \hat{i} \pm \sin(\eta \theta) \hat{j} \right] + \frac{\hbar}{2} \left( \frac{\eta s \Delta_{so}}{E} \right) \hat{k}. \qquad (5)$$

The spin-valley dependent wave function satisfying eq. 4 are given by

$$\left|\psi_{\eta s,\pm k_y}\right\rangle = \frac{1}{\sqrt{1+\left|A_{\eta s}\right|^2}}\begin{pmatrix}1\\A_{\eta s}e^{\pm i\eta\theta}\end{pmatrix}e^{\pm ik_y y+k_{//}x}, \qquad (6)$$

where $A_{\eta s} = \dfrac{E-\eta s\Delta_{so}}{\eta\sqrt{E^2-\Delta_{so}^2}}$. As represented eq.5, we will see that lattice-spin in the z-component depends on real-spin and valley. When $E\to\Delta_{so}$, we get

$$\left\langle\vec{S}_{sp}\right\rangle^z \to \eta s\frac{\hbar}{2}. \qquad (7)$$

This is to say electrons in $k$-valley with real-spin $\uparrow$ and electrons in $k'$-valley with real spin $\downarrow$ acquire lattice-pseudo spin up denoted by $\Uparrow$. Also, electrons in $k$-valley with real-spin $\downarrow$ and electrons in $k'$-valley with real-spin $\uparrow$ acquire lattice-spin up denoted by $\Downarrow$. Hence, we can also say that the currents flowing in strained barrier junction (see Fig.1b) may be considered as carried by two current groups $I_\Uparrow \equiv I_{k\uparrow}+I_{k'\downarrow}$ and $I_\Downarrow = I_{k\downarrow}+I_{k'\uparrow}$. In NM-regions, we get energy-momentum relation as of the form

$$E^2-\Delta_{so}^2 = \left(v_F\hbar k_{//}\right)^2 + \left(v_F\hbar k_y\right)^2. \qquad (8)$$

In St-region, we get

$$\left(E+\mu\right)^2 - \left(\Delta_z+\eta\Delta_\Omega+\eta s\Delta_{so}\right)^2 = \left(v_F\hbar k_{//}+\eta\delta\right)^2 + \left(v_F\hbar q_y\right)^2, \qquad (9)$$

where $\vec{q}=\left\langle k_{//},q_y\right\rangle$ is wave vector component of electrons in St-region. Using eqs. 8 and 9, we may get the energy gap formula of the form $E_{g,\eta s}=2\left|\Delta_z+\eta\Delta_\Omega+\eta s\Delta_{so}\right|$. In the first case, when $\Delta_z=\Delta_{so}$ and $\Delta_\Omega=0$, the lattice-spin-dependent band structure is $E_{g,k(k')\Uparrow}=4\left|\Delta_{so}\right|$ but $E_{g,k(k')\Downarrow}=0$, as shown in Fig.1c. In the second case, when $\Delta_z=0$ and $\Delta_\Omega=\Delta_{so}$, the real-spin-dependent band structure is $E_{g,k(k')\uparrow}=4\left|\Delta_{so}\right|$ but $E_{g,k(k')\downarrow}=0$, as shown in Fig.1d. This property is very significant for studying the possibility to use electric field and circularly polarized photon to completely control lattice-spin and real-spin in silicene-based strained junction.

## 3. Scattering process

In this section, the scattering process in the system is modeled based on the Hamiltonians in eqs. 3 and 4 given in section 2. The wave function of electron in each region is defined as

$$\psi(y<0)=\begin{pmatrix}1\\A_{\eta s}e^{i\eta\theta}\end{pmatrix}e^{ik_y y+k_{//}x}+r_{\eta s}\begin{pmatrix}1\\A_{\eta s}e^{-i\eta\theta}\end{pmatrix}e^{-ik_y y+k_{//}x},$$

$$\psi(0 \leq y < L) = a_{\eta s}\begin{pmatrix} 1 \\ B_{\eta s}e^{i\eta\phi} \end{pmatrix}e^{iq_y y + k_{//}x} + b_{\eta s}\begin{pmatrix} 1 \\ B_{\eta s}e^{-i\eta\phi} \end{pmatrix}e^{-iq_y y + k_{//}x}$$

and

$$\psi(L \leq y) = t_{\eta s}\begin{pmatrix} 1 \\ A_{\eta s}e^{i\eta\theta} \end{pmatrix}e^{ik_y y + k_{//}x},$$

(10)

where $B_{\eta s} = \dfrac{E + \mu - (\Delta_z + \eta\Delta_\Omega + \eta s\Delta_{so})}{\eta\sqrt{(E + \mu)^2 - (\Delta_z + \eta\Delta_\Omega + \eta s\Delta_{so})^2}}$. The angle $\phi$ in the St-region may be determined via the formula

$$v_F\hbar k_{//} + \eta\delta = \sqrt{(E + \mu)^2 - (\Delta_z + \eta\Delta_\Omega + \eta s\Delta_{so})^2}\cos\phi.$$ (11)

The coefficients $a_{\eta s}$ and $b_{\eta s}$ are denoted as the amplitudes of plane wave scattering in the St-region. The reflected and transmitted coefficients are $r_{\eta s}$ and $t_{\eta s}$, respectively. They may be determined using the boundary conditions at the interfaces $\psi(y = 0^+) = \psi(y = 0^-)$ and $\psi(y = L^+) = \psi(y = L^-)$. By doing this, the transmitted coefficient may be obtained as

$$t_{\eta s} = \frac{A_{\eta s}B_{\eta s}\left(e^{2i\eta\theta} - 1\right)\left(e^{2i\eta\phi} - 1\right)e^{iq_y L - ik_y L}}{\Pi_{\eta s}},$$

where

$$\Pi_{\eta s} = A_{\eta s}B_{\eta s}e^{2i\eta(\theta+\phi)} + \left(A_{\eta s}^2 + B_{\eta s}^2\right)e^{i\eta(\theta+\phi)}\left(e^{2iq_y L} - 1\right) - A_{\eta s}B_{\eta s}\left(e^{2i\left(q_y L + \eta\theta\right)} + e^{2i\left(q_y L + \eta\phi\right)} - 1\right).$$

(12)

## 4. Real and lattice-spin-dependent conductances

As discussed in section 2, the real and lattice-spin-dependent transmissions in $k$ and $k'$ valleys may be defined as of the forms

$$T_{k\Uparrow} = T_{k\uparrow} = |t_{k\uparrow}|^2 \ , \ T_{k\Downarrow} = T_{k\downarrow} = |t_{k\downarrow}|^2$$

and

$$T_{k'\Uparrow} = T_{k'\downarrow} = |t_{k'\downarrow}|^2, \ T_{k'\Downarrow} = T_{k'\uparrow} = |t_{k'\uparrow}|^2 \ .$$

(13)

The real and lattice-spin conductances are related to the transmissions in eq. 13. They are given by using the Landauer formula [35] which integrates transmissions over all incident angle, as given by

$$G_{k\uparrow(\downarrow)} = \frac{1}{2}G_0\int\limits_0^\pi d\theta \sin\theta T_{k\uparrow(\downarrow)} \text{ and } G_{k'\uparrow(\downarrow)} = \frac{1}{2}G_0\int\limits_0^\pi d\theta \sin\theta T_{k'\uparrow(\downarrow)},$$

$$G_{k\Uparrow(\Downarrow)} = \frac{1}{2}G_0\int\limits_0^\pi d\theta \sin\theta T_{k\Uparrow(\Downarrow)} \text{ and } G_{k'\Uparrow(\Downarrow)} = \frac{1}{2}G_0\int\limits_0^\pi d\theta \sin\theta T_{k'\Uparrow(\Downarrow)},$$

$$(14)$$

where $G_0 = \frac{4e^2}{h}N(E)$ is a conductance of pure silicene with $N(E) = \frac{w}{\pi\hbar v_F}\sqrt{E^2 - \Delta_{so}^2}$ being the density of state for non-impurity silicene. The width of the junction is denoted as $w$.
Thus, the total real and lattice-spin conductances take the forms

$$G_{\uparrow(\downarrow)} = G_{k\uparrow(\downarrow)} + G_{k'\uparrow(\downarrow)} \text{ and } G_{\Uparrow(\Downarrow)} = G_{k\Uparrow(\Downarrow)} + G_{k'\Uparrow(\Downarrow)}, \qquad (15)$$

respectively. The net conductance in the junction is given by $G = G_\Uparrow + G_\Downarrow = G_\uparrow + G_\downarrow$. The
lattice-spin currents are related to lattice-spin conductances $I_\Uparrow \approx G_\Uparrow$ and $I_\Downarrow \approx G_\Downarrow$, where the
net current in the junction may be given by $I = I_\Uparrow + I_\Downarrow = I_\uparrow + I_\downarrow$.

## 5. Results and discussion

In our numerical study, we focus on three cases: (i) the case that yields perfect strain
control of pure lattice-spin, (ii) the case that yields perfect strain control of pure real-spin, and
(ii) the case that yields current carried by three groups of electron species.

## (i) Perfect strain control of pure lattice-spin by applying electric field

In this case we apply perpendicular electric field without circularly polarized photon
onto the barrier to set to be $\Delta_z = \Delta_{so}$ and $\Delta_\Omega = 0$. This condition may give rise to energy gap
of $E_{g,k(k')\Uparrow} = 4\Delta_{so}$ and $E_{g,k(k')\Downarrow} = 0$. The barrier turns into insulator for electrons with lattice-
spin $\Uparrow$ (or massive fermions), while it turns into conductor for electron with lattice-spin $\Downarrow$ (or
massles fermions). Thus, strong different transport property between lattice-spin up and down
occurs when taking the system into this condition. For honeycomb atomic structure, it has
been known that small perturbed strain is proportional to magnitude of $\delta$ [36, 39]. Strain in
silicene can be either positive or negative. Firstly, we would like to show the resonant peak
under varying strain. In Figs.2a and 2b, it is found that this junction gives rise to lattice-spin
polarized currents $(G_{k\Uparrow} = G_{k'\Uparrow}) \neq (G_{k\Downarrow} = G_{k'\Downarrow})$. We also show that the conductance peaks of
$G_{k(k')\Uparrow}$ and $G_{k(k')\Downarrow}$ about $\delta = 0$ may occur when a suitable value of chemical potential is

chosen (see Fig.2b). The presence of conductance peak is similar to that in gapped graphene strained junction [41], when energy approaches spin-orbit interaction $E \rightarrow \Delta_{so}$ in the NM regions. The energy gap in pure graphene is due only to sublattice symmetry breaking, unlike silicene. The strain filtering effect in silicene is lattice-spin-dependent, generated from spin-valley Dirac mass terms of the form

$$m_{\eta s} = (\Delta_z + \eta \Delta_\Omega + \eta s \Delta_{so}) / v_F^2 . \qquad (16)$$

Resonance of lattice-spin conductances for $\Uparrow$ and $\Downarrow$ appearing at different $\delta$ is due to the condition that $q_{y\Uparrow}(\delta) \neq q_{y\Downarrow}(\delta)$ inside the barrier, which is a suitable bound state at different strain values. From eq.9, it may be approximated in this case to get

$$q_{y\Uparrow(\Downarrow)} \cong \sqrt{(E+\mu)^2 - (\Delta_{so} + (-)\Delta_{so})^2 - (\delta)^2} / v_F \hbar .$$

This effect leads to completely controllable lattice pseudospin currents by strain in silicene-based system. The peaks appear due to $q_{y\Uparrow} L \neq q_{y\Downarrow} L$, because lattice-spin up and down are bound states at different $\delta$.

We next investigate the translation of conductance peaks about $\delta = 0$ at $L = 100$ nm by varying chemical potential $\mu$ in the strained barrier (see Figs. 3a and 3b). We find that from $\mu = -18$ meV to $-7.4$ meV, the junction yields single-resonant conductance peak which may be shifted by varying $\mu$ along the $\delta-$axis for $0 \leq \delta$, while for $\mu > -7.4$ meV the double peaks would appear [41]. The conductance vanished for large $\delta$ may be described with decaying wave function to get $G_{\Uparrow(\Downarrow)} \propto e^{-L\sqrt{\delta^2 - (E+\mu)^2 + (\Delta_{so} + (-)\Delta_{so})^2}/v_F \hbar}$. The range of $\mu$ that yields single- conductance peak would be applicable for a strain filtering device.

In Figs.4a and 4b, it is shown that for large barrier thickness $L = 285$ nm, strain filtering effect occurs only for $\Downarrow$-current from $\mu = -3$ meV to $-0.25$ meV. This is because $\Uparrow$-current has been suppressed due to the fact that electron acquiring energy gap of $E_{g\Uparrow} = 4\Delta_{so}$ inside the barrier, while electron with lattice-spin $\Downarrow$ carrying energy gap of $E_{g\Downarrow} = 0$. This leads to strong insulator only for $\Uparrow$-electron.

Based on the result above, it is pointed out that our model may be applicable for devices that control lattice-spin currents by strain (see Fig.5). For small barrier thickness $L = 100$ nm (see Fig.5a), the net conductance may give rise to resonant-conductance peaks for $\Uparrow$- and $\Downarrow$- currents at strain of $\pm\delta_{max\Uparrow}$ and $\pm\delta_{max\Downarrow}$, respectively. Lattice-pseudospin currents are nearly completely filtered and controlled by strain. Switching of current in the junction from $\Uparrow$ -current to $\Downarrow$ - current by strain is possible for small barrier thickness. In contrast to the small barrier thickness, the net current is almost governed only by $\Downarrow$ -current for large barrier thickness $L = 285 \, nm$ (see Fig.5b). Resonant conductance peak in the $\delta$ -axis is associated only with $\Downarrow$ -electron, yielding strain $\pm\delta_{max}$ to create $\Downarrow$ -resonant peaks. Finally, the $\delta_{max}$ as a function of $\mu$ is plotted for large barrier thickness (see Fig.6). This is the characteristic of the junction that external strain may be filtered since the current is flow strongly only with specific strain values. This result is similar to that in graphene system [41], which shows strain filtering effect. However, it is not pure lattice-spin current in graphene.

**(ii) Perfect strain control of pure real-spin by applying circularly polarized photon**

In this case we apply circularly polarized photon onto the barrier without perpendicular electric field to set $\Delta_z = 0$ and $\Delta_\Omega = \Delta_{so}$ . In contrast to the first case, this condition may give rise to energy gap of $E_{g,k(k')\uparrow} = 4\Delta_{so}$ and $E_{g,k(k')\downarrow} = 0$ . In this case, the behavior of real-spin conductance under applied strain is the same as given in Figs 2-6, because

$$E_{g,k(k')\uparrow}(\Delta_z = 0, \Delta_\Omega = \Delta_{so}) = E_{g,k(k')\Uparrow}(\Delta_z = \Delta_{so}, \Delta_\Omega = 0) = 4\Delta_{so}$$

and $\qquad E_{g,k(k')\downarrow}(\Delta_z = 0, \Delta_\Omega = \Delta_{so}) = E_{g,k(k')\Downarrow}(\Delta_z = \Delta_{so}, \Delta_\Omega = 0) = 0$ .

Therefore we can use Figs 2-6 to describe real-spin currents in this case by replacing $\Uparrow$ with $\uparrow$ and $\Downarrow$ with $\uparrow$ . It can be thus concluded that the junction may give rise to perfect strain control of real-spin by circularly polarized photon irradiated onto the barrier. This may be applicable for real-spin-based strain sensor.

**(iii) Current carried by three groups of electron species**

In this case, we apply both perpendicular electric field and circularly polarized light into the barrier $\Delta_z = \Delta_\Omega = \Delta_{so}$ to get spin-valley-dependent energy gap and Dirac mass in the St-barrier to get $E_{g,k\uparrow} = 6\Delta_{so}$ and $E_{g,k\downarrow} = E_{g,k'\uparrow} = E_{g,k'\downarrow} = 2\Delta_{so}$ . For small barrier thickness $L = 100nm$ , the two conductance peaks appear as a function of strain (see Fig.7a). One peak is governed by $G_{k\uparrow}$ the other is governed by $G_{k\downarrow}, G_{k'\uparrow}$ and $G_{k'\downarrow}$ . It is seen that the peaks which appear at the same strain, carry three species of electrons, giving rise to $G_{k\downarrow} < G_{k'\uparrow} = G_{k'\downarrow}$ despite having the same energy gap. This behavior may be described by the role of lattice-spin torque [38,42]. The direction of lattice-spin can be determined via the sign of its mass (see eqs. 5 and 7). When electrons in k-valley with $\downarrow$ change its lattice-spin from $\Uparrow$ in the NM-regions to $\Downarrow$ in the St-region, lattice-spin torque occurs to reduce its transmission while there is no lattice-spin torque for electrons of $k' \uparrow (\downarrow)$. This is the result why $G_{k\downarrow}$ is smaller than $G_{k'\uparrow} = G_{k'\downarrow}$, despite having the same energy gap magnitude (see table I). In Fig.7b, it is also found that when L increase up to $L = 285\,nm$, $G_{k\downarrow}$ is almost zero. The current in the junction is completely governed by three electron groups, to get $G_{k\uparrow} < G_{k'\uparrow} = G_{k'\downarrow}$. In Fig. 8a, the translation of the net conductance peak governed by three electron species, $G \cong G_{k\downarrow} + G_{k'\uparrow} + G_{k'\downarrow}$ or $G \cong G_{k\Downarrow} + G_{k'\Downarrow} + G_{k'\Uparrow}$, by varying $-4\,meV < \mu < -1.6\,meV$, is investigated. Plot of $\delta_{max}$ as a function of $\mu$ in Fig. 8b is to show the controllable strain sensor by gated potential. Three electron groups are allowed to transport at $\delta = \pm\delta_{max}$ and $\delta_{max}$ depends on chemical potential in the St-region.

## 6. Summary and conclusion

We have investigated spin-valley transport in silicene-based strained barrier junction where chemical potential, electric field and circularly polarized light are applied into the strained barrier. We focus on investigation of real- and lattice-spin currents controlled by strain. As a result, (i) in the case of applying only electric field ($\Delta_z = \Delta_{so}$ and $\Delta_\Omega = 0$), it has been found that, for small strain-barrier thickness, resonant conductance peaks of $I_\Uparrow$ and $I_\Downarrow$, are almost perfectly split in strain-axis. This leads to perfect strain control of lattice-pseudospin currents. The resonant conductance peak is solely associated with $I_\Downarrow$ to get lattice-spin-based strain filtering effect in the case of large barrier thickness. (ii) In the case of applying only circularly polarized photon ($\Delta_z = 0$ and $\Delta_\Omega = \Delta_{so}$), conductance peaks of $I_\uparrow$ and $I_\downarrow$ are found to be almost perfectly split in strain-axis. This leads to perfect strain control of real-spin currents. In this case, the conductance peak is solely associated only with $I_\downarrow$ to get real-spin-based strain filtering effect for large barrier thickness. It may be said that electric field may generate strain control of lattice-spin currents while circularly polarized light may generate real-spin controlled by strain. Interestingly, (iii) in the case of applying both electric field and circularly polarized photon to be $\Delta_z = \Delta_{so} = \Delta_\Omega$, strain filtering effect still occurs but the current is not due to pure lattice- or real- spin currents. The conductance as a function of strain gives rise to two conductance peaks that one is governed by $I_{k\uparrow}$ while the other is governed by three group of electron species $I_{k\downarrow}, I_{k'\uparrow}$, and $I_{k'\downarrow}$. This behavior is considered as surprising and interesting. Strain filtering effect in a silicene-based junction is quite different from that in graphene junction [41]. Our work has given interesting results for application of silicene as an electro-mechanical device and spin-valleytronic applications.

**Acknowledgments**

B. Soodchomshom acknowledges Kasetsart University Research and Development Institute (KURDI) and Thailand Research Fund (TRF) under Grant. No. RSA5980058. This research is supported in path by the Graduate Program Scholarship from the Graduate School, Kasetsart University.

**References**

[1] L. C. Lew Yan Voon, J. Zhu, U. Schwingenschlögl, Appl. Phys. Rev. **3** (2016) 040802

[2] J. Zhao et al., Progress in Materials Science **83** (2016) 24

[3] A. Kara et al., Surface Science Reports **67** (2012) 1

[4] P. Vogt et al., Phys. Rev. Letts. **108** (2012)155501

[5] L. Tao et al., Nature nanotechnology **10** (2015) 227

[6] Q. Ru-Ge et al., Chin. Phys. B **24** (2015) 088105

[7] J. Prasongkit et al., J. Phys. Chem. C **119** (2015) 16934

[8] B. Soodchomshom, Journal of Applied Physics **115** (2014) 023706

[9] N. D. Drummond, V. Zólyomi, V. I. Fal'ko Phys. Rev. B **85**(2012) 075423

[10] M. Ezawa, Phys. Rev. Lett. **110** (2013) 026603

[11] M. Ezawa, Phys. Rev. Lett. **109** (2012) 055502

[12] C.-C. Liu, H. Jiang, Y. Yao , Phys. Rev. B **84** (2011) 195430

[13] M. Ezawa, Phys. Rev. B 87 (2013) 155415

[14] X. Zhai, G. Jin, Journal of Physics: Condensed Matter 28 (2016) 355002

[15] T. Yokoyama, New J. Phys. **16** (2014) 085005

[16] X. Zhai et al., Applied Physics Letters **109** (2016) 122404

[17] T. Yokoyama, Phys. Rev. B **87** (2013) 241409(R)

[18] M. Tahir et al., Appl. Phys. Lett. **102** (2013) 162412

[19] Y. Mohammadi,B. A. Nia, Superlattices and Microstructures **96** (2016) 259

[20] T. Hua et al., Solid State Communications **244** (2016) 43

[21] Q-X. Pei et al., J. Appl. Phys. **115** (2014) 023519

[22] Y-C. Fan, T-H. Fang, T-H. Chen, Nanomaterials **6** (2016) 120

[23] Y. D. Kuang et al., Nanoscale **8** (2016) 3760

[24] Qin et al. Nanoscale Research Letters **9** (2014) 521

[25] J-A. Yan et al., Phys. Rev. B **91** (2015) 245403

[26] C. Yesilyurt et al., Applied Physics Express **8** (2015) 105201

[27] W. Sa-Ke, W. Jun, Chin. Phys. B **24** (2015) 037202

[28] Q-X. Pei et al., J. Appl. Phys. **114**(2013)033526

[29] P. Chantngarm, K. Yamada, B. Soodchomshom, Superlattices and Microstructures **94** ( 2016)13

[30] D. Huertas-Hernando et al., Phys. Rev. B **74** (2006) 155426

[31] H.Min et al., Phys. Rev. B **74**(2006)165310

[32] S. Konschuh, M. Gmitra, J. Fabian, Phys. Rev. B **82**(2010) 245412

[33] M. Ezawa, Phys. Rev. B **88** (2013) 161406 (R)

[34] P. Chantngarm, K. Yamada, B. Soodchomshom, Journal of Magnetism and Magnetic materials **429** (2017)16

[35] R. Landauer, IBM Journal of Research and Development **1** (1957) 223.

[36] Vitor M. Pereira et al., Phys. Rev. B **80**(2009) 045401

[37] M. Ezawa , J. Phys. Soc. Jpn. **84** (2015)121003

[38] K. Jatiyanon, B. Soodchomshom, Physica E **80** (2016)120

[39] Vitor M. Pereira, A. H. Castro Neto, Phys. Rev. Lett. **103**(2009) 046801

[40] J. Zhou et al., Nano Res. **9** (2016) 1578

[41] T. Chethanom, R. Jongchotinon, B. Soodchomshom, Superlattices and Microstructures 85 (2015) 716

[42] L. Majidi, M. Zareyan, Phys. Rev. B **83**(2011) 115422

| Case III | $m_{k\uparrow}$ | $m_{k\downarrow}$ | $m_{k'\uparrow}$ | $m_{k'\downarrow}$ |
|---|---|---|---|---|
| NM-region | $+\Delta_{so}/v_F^2$ | $-\Delta_{so}/v_F^2$ | $-\Delta_{so}/v_F^2$ | $+\Delta_{so}/v_F^2$ |
| St-region | $+3\Delta_{so}/v_F^2$ | $+\Delta_{so}/v_F^2$ | $-\Delta_{so}/v_F^2$ | $+\Delta_{so}/v_F^2$ |
| Lattice-spin torque | Yes (not reverse direction) | Yes (reverse direction) | No | No |

**Table I**: Illustration of why $G_{k\downarrow}$ is smaller than $G_{k'\uparrow} = G_{k'\downarrow}$, although they have the same energy gap, in the case iii. Small lattice-spin torque for electron in state $k\uparrow$, larger lattice-spin torque for electron in state $k\downarrow$ and no lattice-spin torque for electron in state $k'\downarrow(\uparrow)$. Large lattice-spin torque may reduce transmission, despite the same magnitude of gap.

**Figure captions**

**Figure 1** Schematic illustrations of (a) atomic geometry of silicene including 3-nerest neighbor and 6-next nearest neighbor atoms where uniaxial strain applied into the armchair direction, (b) a NM/St/NM junction where strain, chemical potential, circularly polarized light and electric field are applied into the St region (c) lattice-pseudospin-dependent dispersion in the St region for the case of $\Delta_{so} = \Delta_z = eE_zD$ and $\Delta_\Omega = 0$, and (d) real-spin-dependent dispersion in the St region for the case of $\Delta_z = 0$ and $\Delta_\Omega = \Delta_{so}$. We take $\Uparrow(\Downarrow)$ for lattice-spin up (down), and $\uparrow(\downarrow)$ stands for real-spin up (down).

**Figure 2** Plot of lattice-pseudospin conductance as a function of strain-induced perturbed hopping energy $\delta$ (a) for $\mu = 0$ and (b) for $\mu = -12\,\text{meV}$. The conductance peaks of lattice-pseudo spin up and down are found at different strain.

**Figure 3** Plot of (a) lattice-pseudospin down conductance and (b) lattice-pseudospin up conductance, as a function of strain-induced perturbed hopping energy $\delta$ for $L = 100\,\text{nm}$. The conductance peaks about $\delta = 0$ are found for both lattice-pseudospin up and down. The positions of the peaks are tunable by chemical potential.

**Figure 4** Plot of (a) lattice-pseudospin down conductance and (b) lattice-pseudospin up conductance, as a function of strain-induced perturbed hopping energy $\delta$ for $L = 285\,\text{nm}$. The conductance peaks about $\delta = 0$ are found only for lattice-pseudospin down, while the lattice pseudospin up currents are very small.

**Figure 5** Both lattice pseudospin up and down conductance peaks under varying strain found for small barrier thickness (a) $L = 100\text{nm}$, while pure lattice pseudospin down conductance peak occurs for large barrier thickness (b) $L = 285\text{nm}$.

**Figure 6** Plot of strain $\delta_{max}$ which yields a single maximum conductance governed by $I_\Downarrow = I_{k\Downarrow} + I_{k'\Downarrow}$ under varying $\mu$. This leads to **tunable strain filter by pure lattice-pseudospin currents.**

**Figure 7** Spin-valley dependent conductances in the case of applying both electric field and circularly polarized light $\Delta_{so} = \Delta_Z = \Delta_\Omega$ (a) for $L = 100\text{nm}$, and (b) for $L = 285\text{nm}$.

**Figure 8** Plot of (a) total conductances in the case of applying both electric field and circularly polarized light $\Delta_{so} = \Delta_Z = \Delta_\Omega$ and $L = 285\text{nm}$ for various values of $\mu$. (b) Plot of strain $\delta_{max}$ which yields maximum conductance governed by three species of electron currents $I_{k\downarrow}$, $I_{k'\uparrow}$, and $I_{k'\downarrow}$, under varying $\mu$.

(1a)

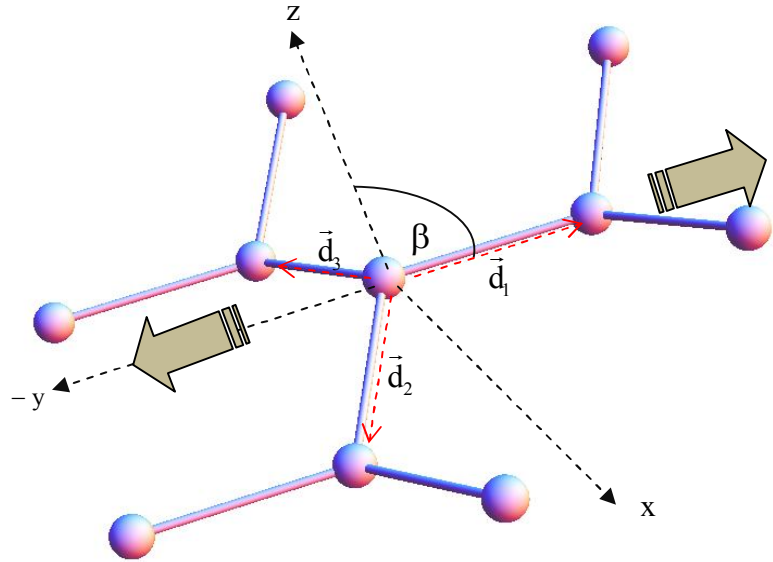

(1b)

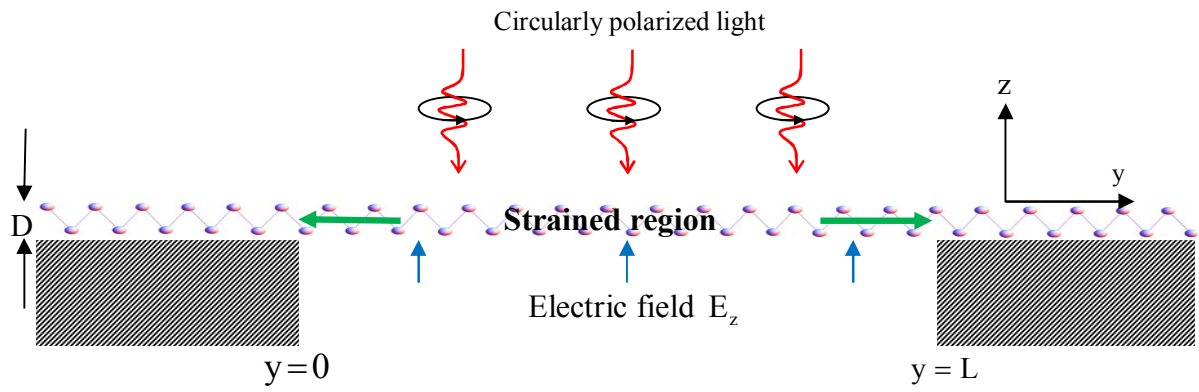

(1c)

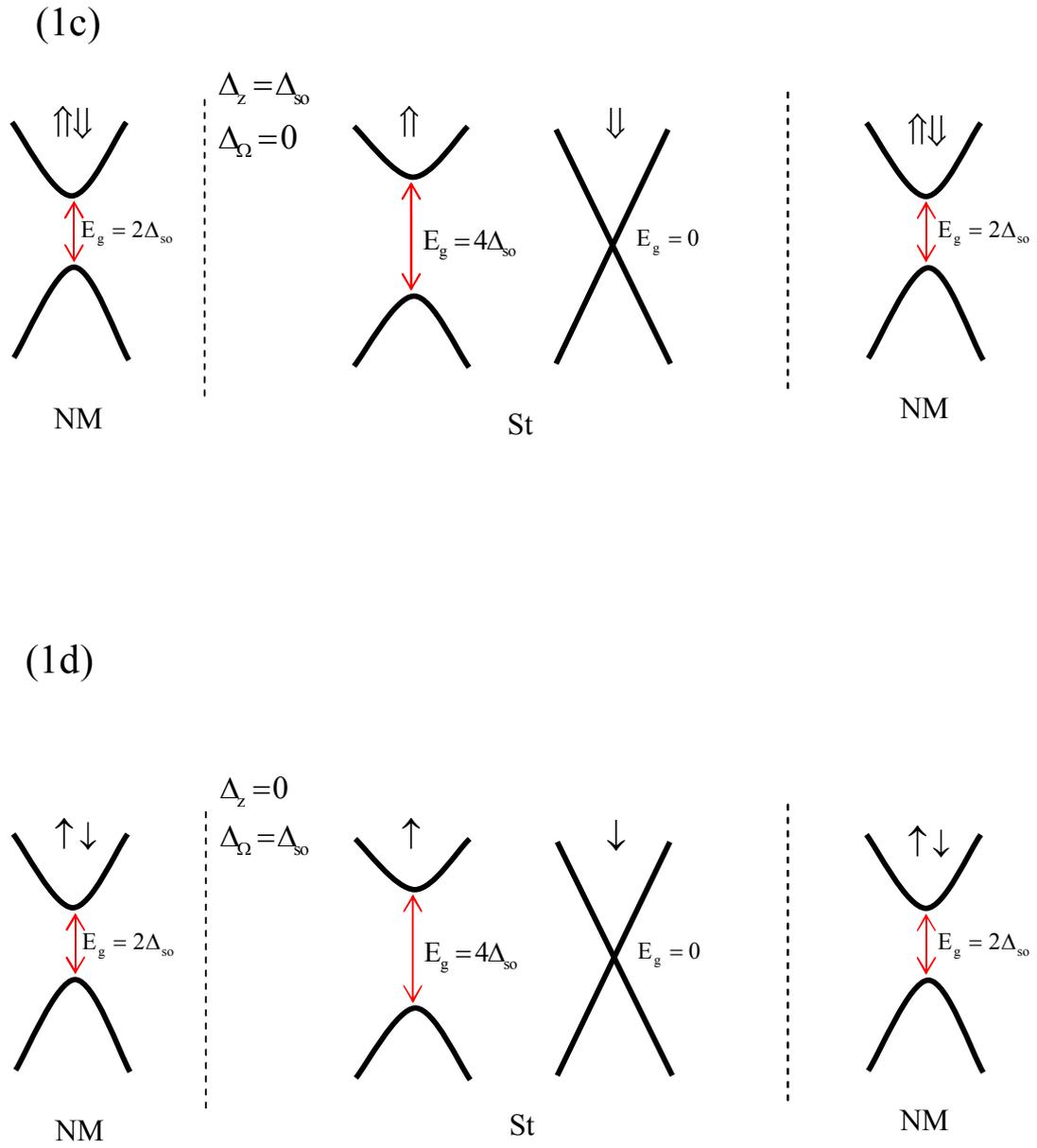

(1d)

**Figure 1**

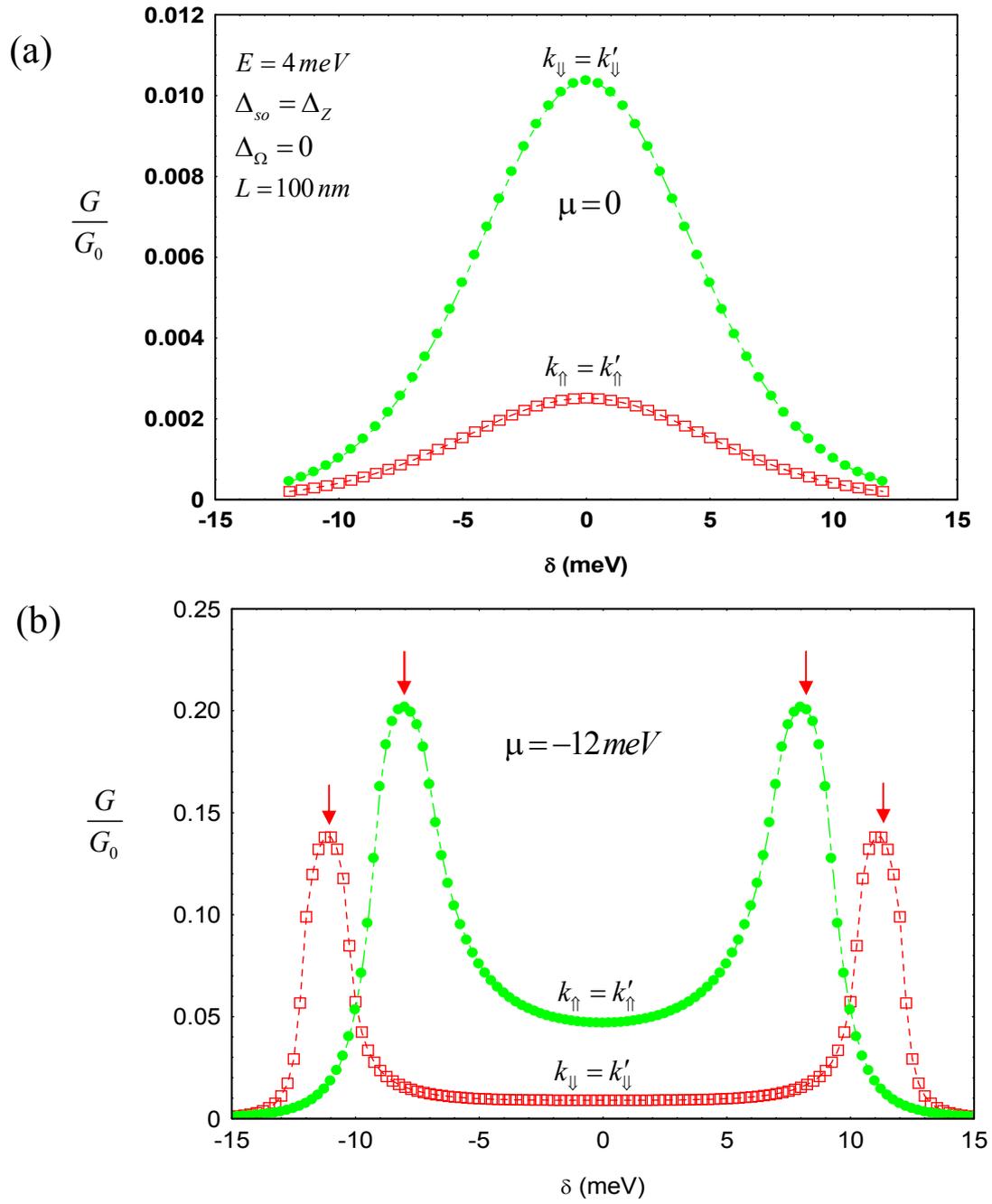

**Figure 2**

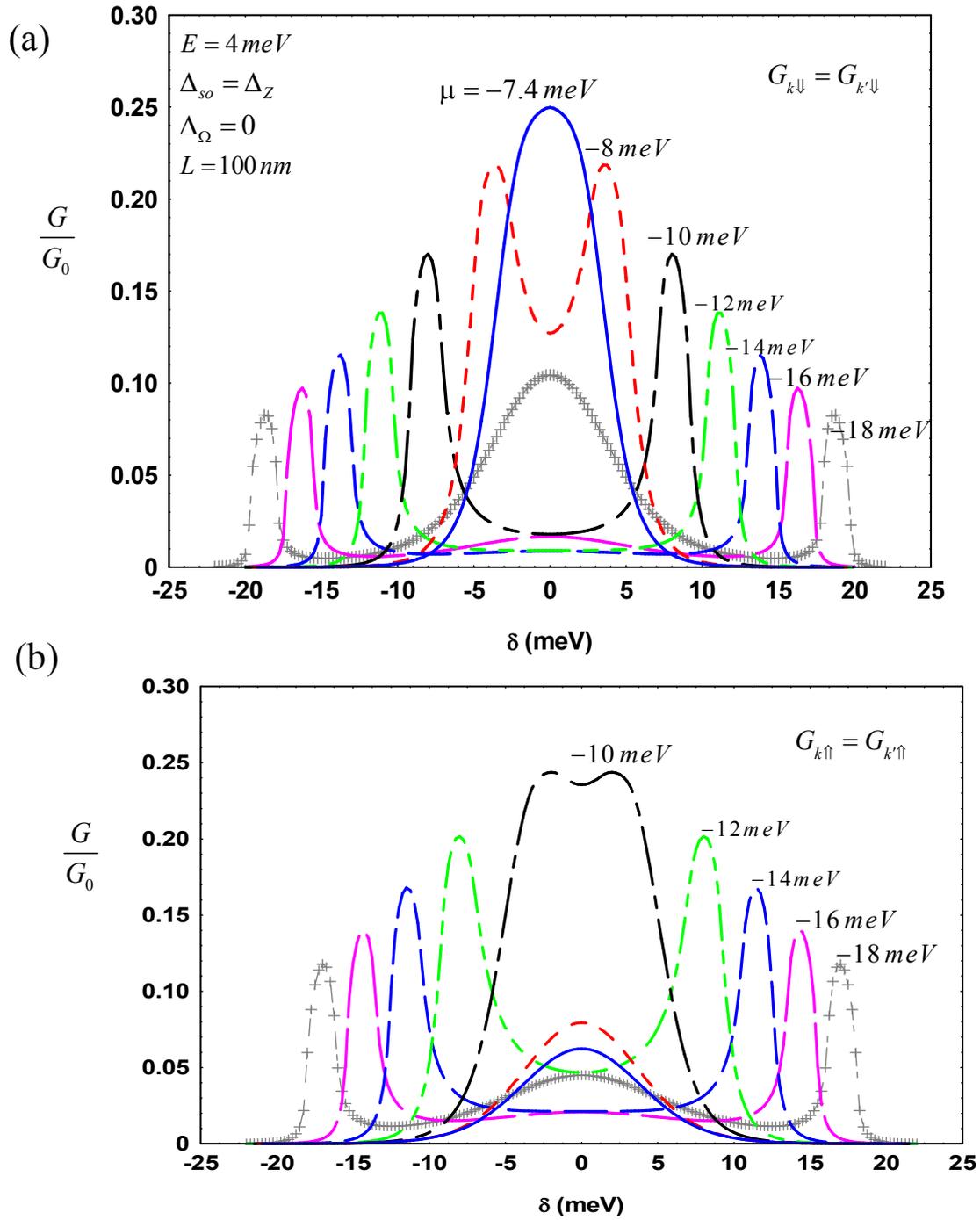

**Figure 3**

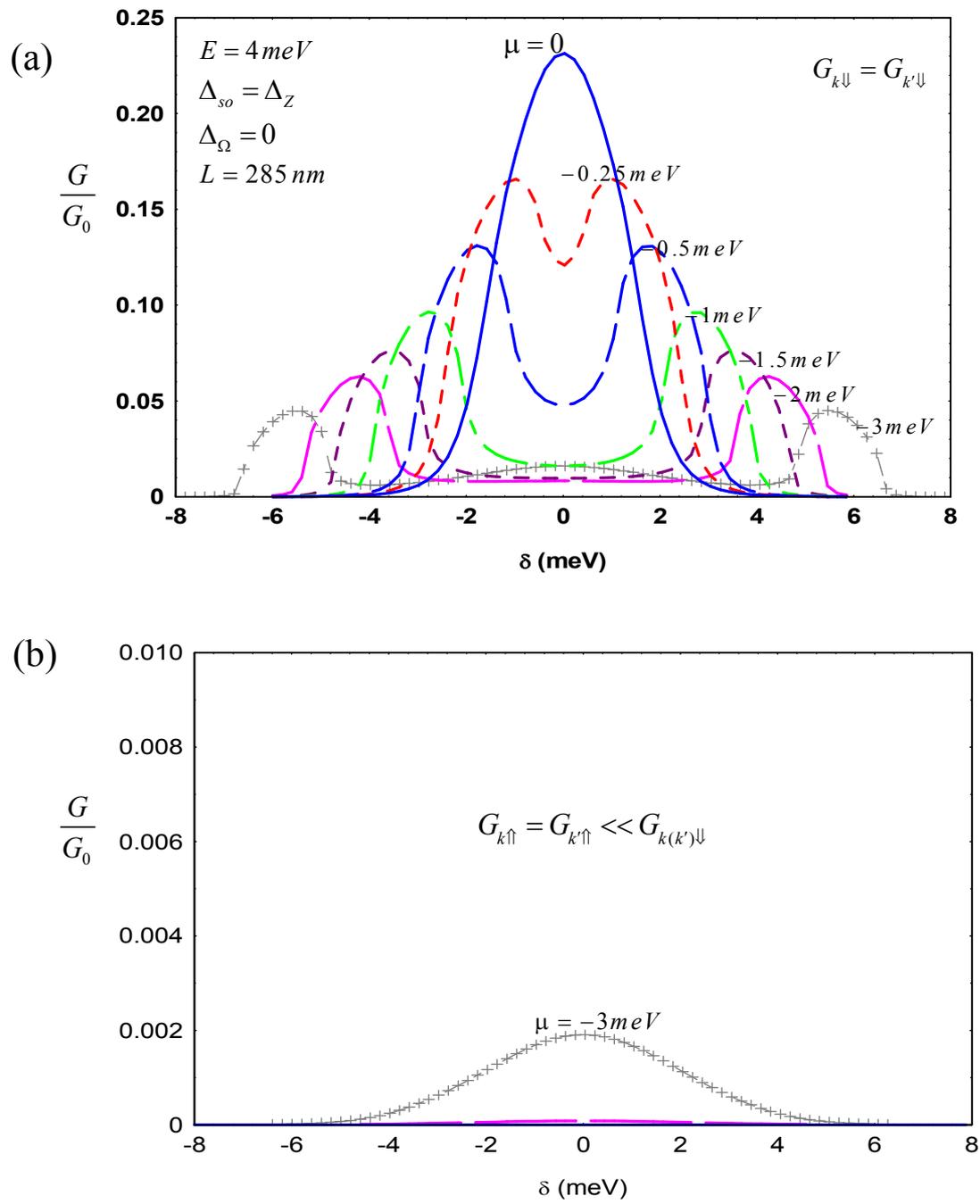

**Figure 4**

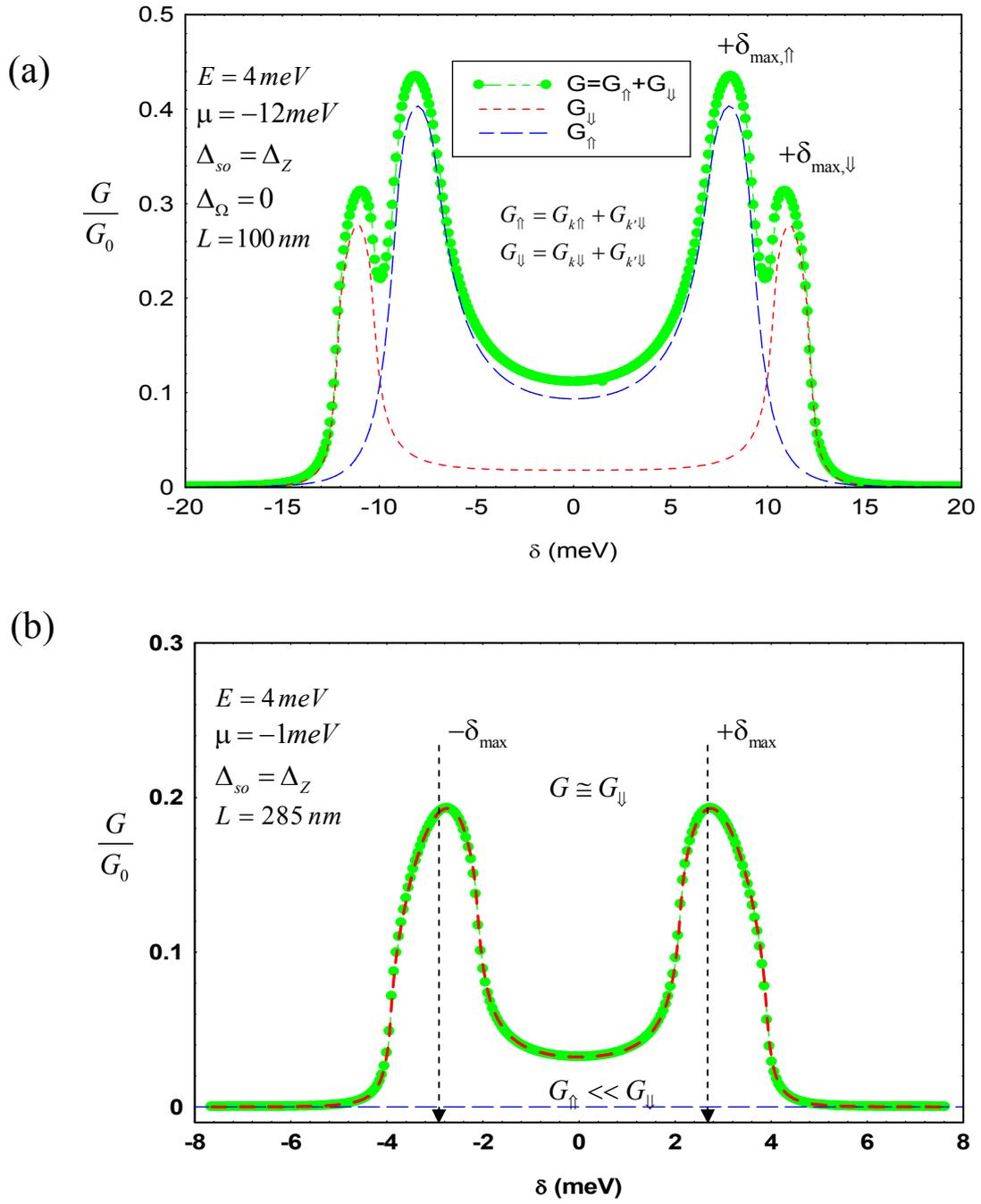

**Figure 5**

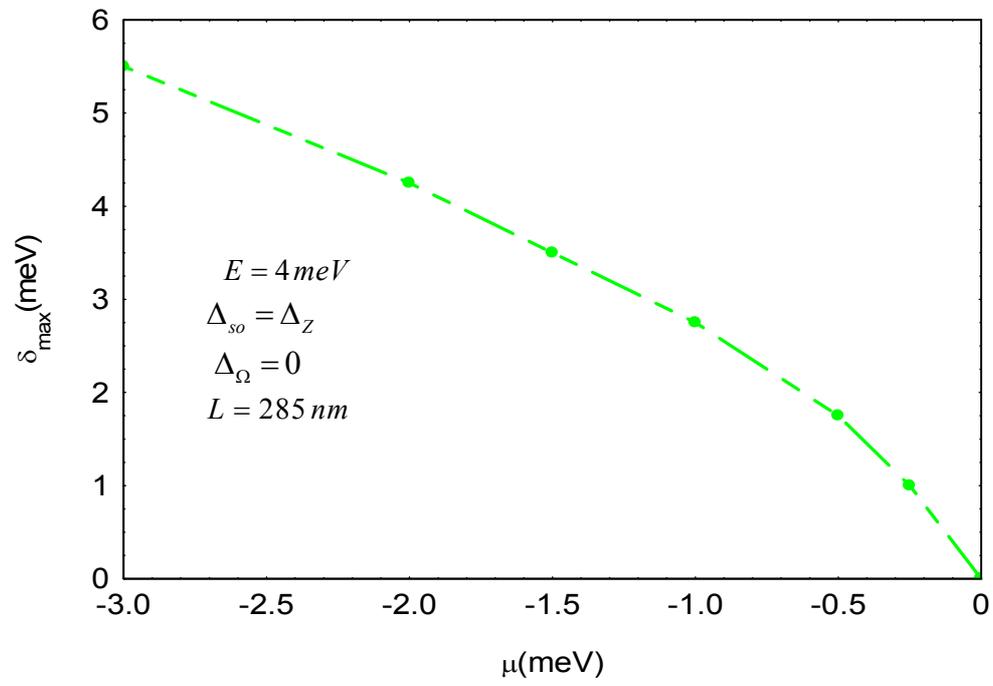

**Figure 6**

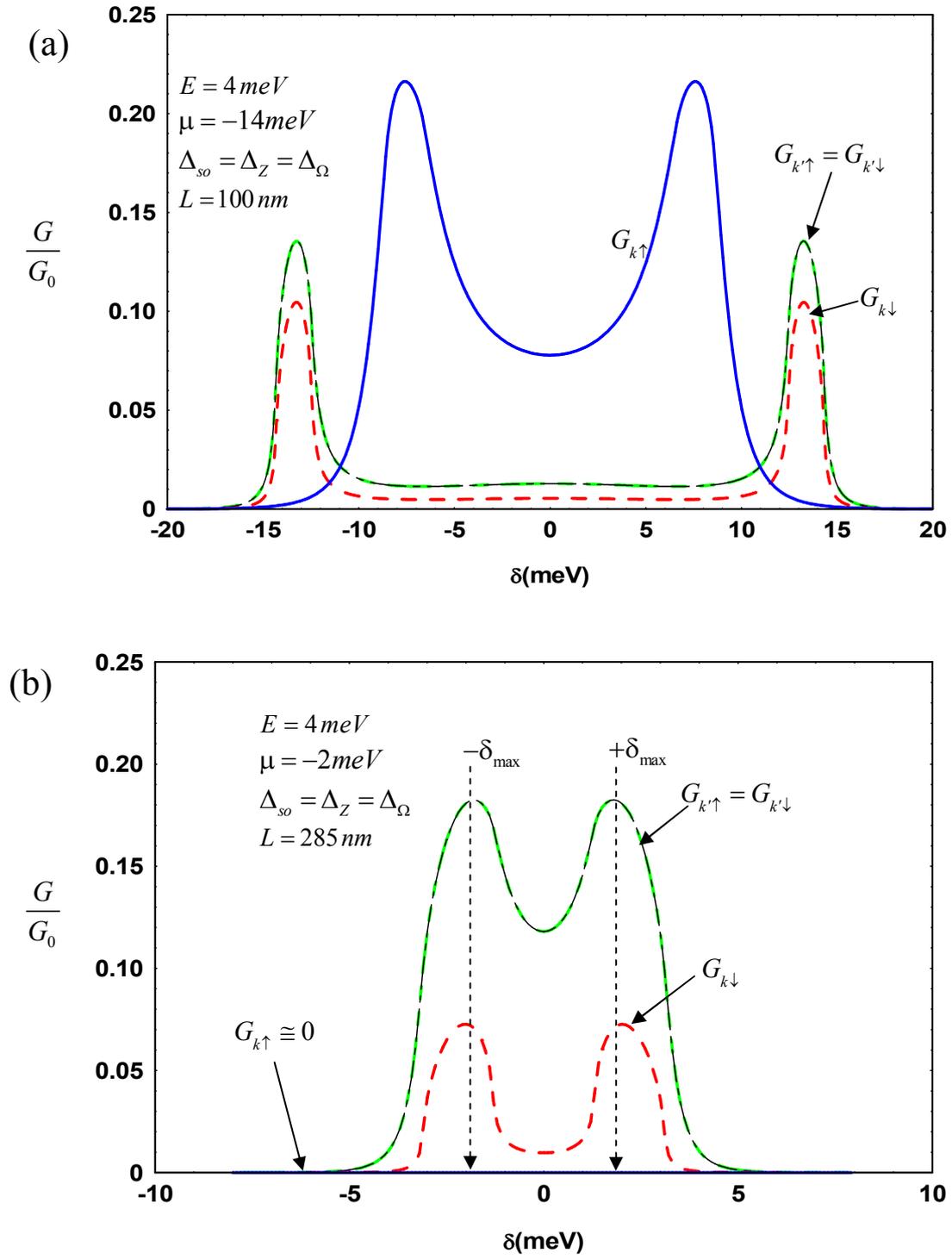

**Figure 7**

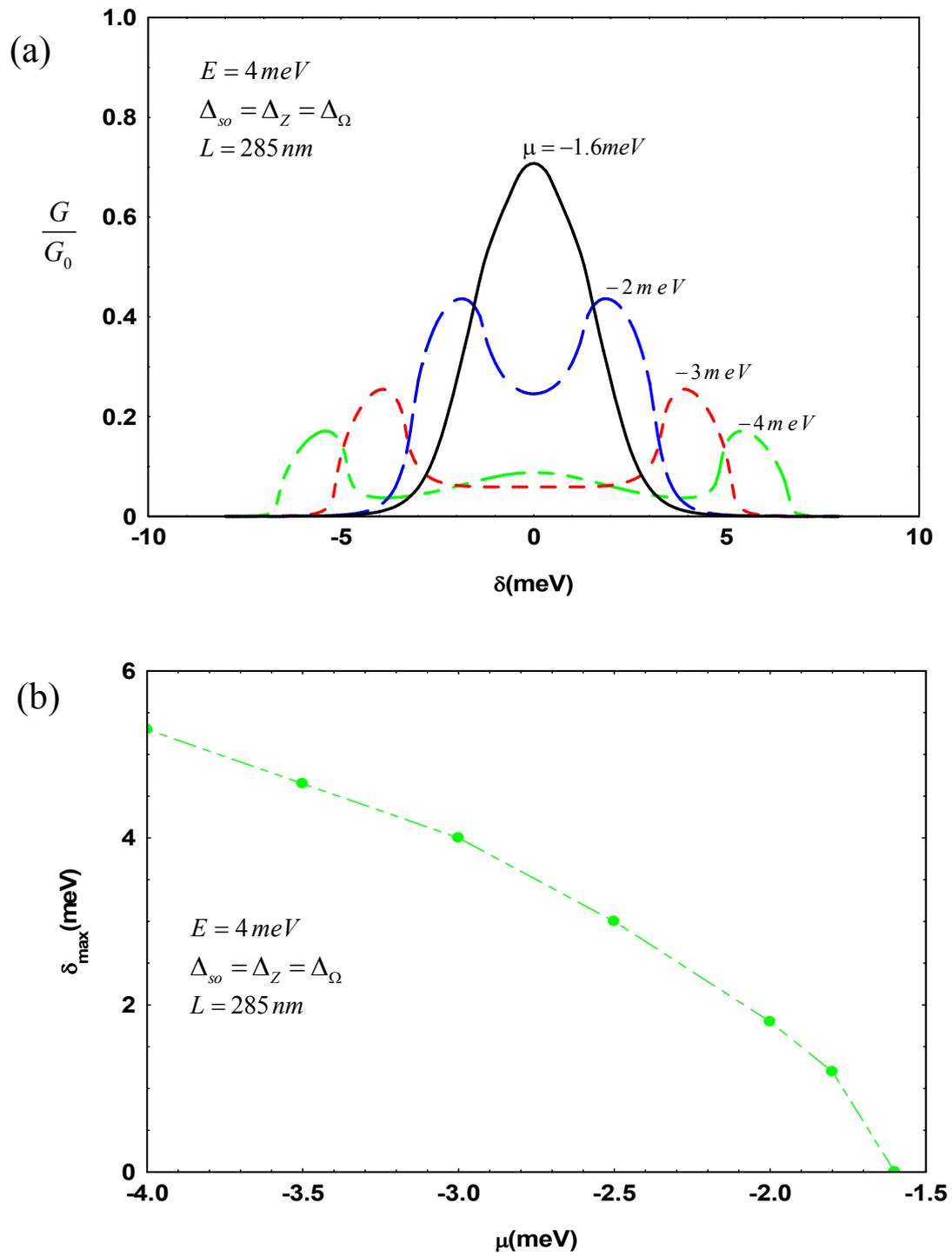

**Figure 8**